# Recent Conceptual Consequences of Loop Quantum Gravity
# Part I: Foundational Aspects

Rainer E. Zimmermann

IAG Philosophische Grundlagenprobleme,
FB 1, UGH, Nora-Platiel-Str.1, D – 34127 Kassel /
Clare Hall, UK – Cambridge CB3 9AL[1] /
Lehrgebiet Philosophie, FB 13 AW, FH,
Lothstr.34, D – 80335 München[2]
e-mail: pd00108@mail.lrz-muenchen.de

## Abstract

Conceptual consequences of recent results in loop quantum gravity are collected and discussed here in view of their implications for a modern philosophy of science which is mainly understood as one that totalizes scientific insight so as to eventually achieve a consistent model of what may be called „fundamental heuristics" on an onto-epistemic background which is part of recently proposed transcendental materialism. This enterprise is being understood as a serious attempt of answering recent appeals to philosophy so as to provide a conceptual foundation for what is going on in modern physics, and of bridging the obvious gap between physics and philosophy. This present first part of the paper deals with foundational aspects of this enterprise, a second part will deal with its holistic aspects.

## Preliminary Statement

In his report on the 13th International Congress on Mathematical Physics (icmp2000) at Imperial College London, Abhay Ashtekar formulates: „In the quantum gravity [contributed] session, while the first two talks were on „standard" mathematical physics topics on the interface of general relativity and quantum physics, the last two were on the interface between quantum gravity, philosophy of science and quantum computing. Unfortunately, this attempt to broaden [sic] and reach out to neighboring field[s] did not succeed; there was a marked difference in the level of precision and emphasis between the two sets of talks." [In: http://gravity.phys.psu.edu/mog/mog16/node13.html] He forgot to tell that he himself as the session's chairman found it necessary at the time to explicitly stress this „reaching out“ to other fields twice, namely at the very beginning of the session and immediately before the first talk of the second half. And this announcement was made in such an apologetic tone that it was clearly signalled to most of the audience that now would come the right moment for a little nap. In fact, the announcement could well be read as some sort of excuse for having been forced to take on these two talks despite being chairman. The reaction was equally obvious and clear, particularly among the younger colleagues who still believe in authority rather than in their own critical abilities. (Letting aside the fact that the available time for the talks was decisively shortened without prior notice and one of the overhead projectors broke down - i.e., that the session was not very well organized anyway.) So this attempt could hardly be called an experiment. Instead, one should call it a self-fulfilling prophecy. The question is whether it is really useful (and serves the purpose) to define everything that is being done by one group of people (let us call it the Pittsburgh group) as an advancement of the desired interaction between physics and philosophy, and to negate alternative approaches without actually discussing them at all. Precision and emphasis are aspects which have to be cleared within a framework of open communication, not by means of self-defined verdicts. At least, so is the custom in European continental philosophy.

[1] Permanent addresses.
[2] Present address.

## Introduction

It has been shown at other occasions [6] that recent results of modern physics can be used to shed some more light onto the foundations of the world, provided the actual task of philosophy is being re-interpreted in terms of a theory which is following up the results of science rather than laying the grounds for the latter, contrary to what the original intention of Aristotelian *prima philosophia* would imply. As it turns out, the interpretation of the main results of present research dealing with aspects of quantum information theory and quantum gravity, respectively, as well as with self-organized criticality, suffices to re-construct a large class of phenomena not only within the field of physics proper, but also within chemistry, biology, and even the social sciences.
As seen under a *philosophical* perspective, it can be shown that the general conceptualization of such a unified view of the world has been prepared on a long line of thought which begins with the Greek Stoá and leads up to the theories of Spinoza, Schelling and Bloch as some of its representatives, eventually showing up in a somewhat modified form in what can be called *transcendental materialism* today. ([6], reference 5) As seen under the *physical* perspective however, it can be demonstrated how a philosophy re-interpreted in the above-mentioned sense can unfold a heuristic potential which is able to produce guidelines of orientation as to deciding about competing concepts in physics. Such a function of philosophy has often been asked for, but has rarely been realized so far when there actually was one of the few occasions to explicitly deal with it in detail. One problem of this is to sort out what would be the appropriate philosophical approach to pertinent questions, in the first place. In the present paper we shall try to contribute somewhat to a further clarification.

This unsatisfactory development of the past has its reasons, of course. In particular, the more recent philosophy of the 20$^{th}$ century, after world war II, has more or less completely suppressed the line of thinking which has been mentioned above, essentially in favour of a self-centring discussion of philosophy itself, taking its label for its contents. This is due to the neo-Kantian effort dating back to the 19$^{th}$ century, to eventually establish a new, formal unity of the sciences, philosophy, the arts, and religion. Although this mainly apologetic enterprise has been subject to failure from the beginning on, it is nevertheless determining European continental (and in particular German) philosophy still today by its two basic components of activities: the history of philosophy on the one hand and the foundations of concepts on the other. Concrete *praxis* then, shows up only in terms of a developmental history of thinking, divorced from any practical relationship to worldliness and empirical existence – being enshrined in a kind of „pure motion of thoughts“ which has lost already sight of the world of everyday life, of the socially mediated and politically active human being. Hence, the history of philosophy is very often mixed up with philosophy proper, the practise of *doing* history with actual philosophizing. And concentrating on the motion of thoughts means to actually forget about what these thoughts actually do reflect at all – thinking all the time, uncorrectly though, that the latter would not be important, once the mechanism of relexion itself would have been uncovered. This view is also supported, though from another perspective, by analytic philosophy which dominates the Anglo-Saxon countries. Here also, the systematization of thinking is more important than that about which thinking actually thinks. Moreover, it is assumed, also uncorrectly, that what is being thought can always be said, thus denying the necessity for any hermeneutic which is surpassing formal logic.

At the same time, this ideological tendency serves also a much more personal interest, namely one which aims at the securing of an isolated field of discourse which is exclusively reserved for the philosophical discourse alone. This has become very obvious in the recently emerged field of „cognitive science“, where each explicitly materialistic model is being rejected from the outset by this „philosophical lobby“, based on the false premise of indicating that this would be nothing than late relics of the mechanical materialism of the 19$^{th}$ century, and a primarily reductionistic effort with the objective of eventually taking over philosophy in merely scientific terms. However, keeping to this false premise means to forget about the fact that a philosophy which can be understood in terms of a dialectical and transcendental materialism visualized within a modern perspective is nothing but the result of critically reflecting recently gained insight in the various fields of science. In other words: only such an approach can be called *practical* in the first place.

But although being oriented with a view to the sciences, philosophy is not at all duplicating the latter's work. Contrary to that, philosophizing means primarily to critically reflect each single sector of the uncovered worldliness with a view to the global interrelationships among all possible sectors, to ask for necessary and sufficient conditions for the main structural aspects of single sectors, and for possible alternatives. In this sense, philosophical reflexion gains more and more a *heuristic* connotation: The unification of the scientifically (necessary) segmented actuality is no self-purpose therefore: On the contrary, the important point is to show, within all what is empirically observable, what *is not more* subject of science.

In the following we discuss the foundation of the physical world in terms of recent progress made in the understanding of the concept of spin networks. (section 1) We try to clarify the concept of „foundation" itself, and come to a conceptual conclusion which is somewhat different from the common utilization of this concept in physics. Subsequently, in an attempt of introducing the notion of downward and upward causation, we deal with aspects of quantum computation, and with the concepts of loops and knots. (sections 2 and 3) The idea is to find a common basis for building and re-tracing a hierarchy of modelling the actuality of the world as a progressive evolution towards increasing complexity. Hence, both the approach of topological quantum field theory (section 4), and the conception of self-organized criticality (section 5), will be helpful in order to describe the explicitly dynamical architecture of this modelling. Smolin's principle of cosmic selection (section 6) will add some more insight into this, if being visualized in a somewhat modified way. The forthcoming second part of this present paper will deal with a possible generalization of this approach trying to reach out into regions beyond physics, hence trying to „climb" the hierarchy of wordly structures.

## 1 Spin Networks & Spin Foams

Recall that essentially, spin networks in the sense of Penrose [1] are coloured graphs, i.e. trivalent graphs with edges labelled by spins. More generally, the edges can be thought of as being labelled by group representations and the vertices as being labelled by intertwining operators. For the case that the respective group is SU(2) and the graph is trivalent, the original form of a spin network emerges. As Baez has shown in some detail for the simplified case of BF theory [2], the full development of the basic ideas about spin networks is comparatively straightforward, if visualized within the context of loop quantum gravity. The two crucial innovations of the latter are first its insistence on a background-free approach to gravity and second that it uses a formulation of Einstein's equations in which – similar to gauge theory in general – parallel transport rather than the metric is important. Spin networks fit generically into this framework. The im-portant result is that the elementary excitations as represented in terms of Hilbert space terminology (in fact, originating from the quantizing of cotangent bundles associated with a suitable space of connections) turn out to be spin network edges of dimension one. [3] This leads straightforwardly to spin foams as a two-dimensional replacement of one-dimensional Feynman diagrams: A spin foam shows up as a two-dimensional complex built from vertices, edges, and polygonal faces. The edges are labelled by intertwining operators, the faces by group representations. The idea of spin foams is to work with path integrals that can be understood as analogues of the afore-mentioned Feynman diagrams. Consequently, a generic slice of a spin foam is a spin network. The motivation for this comes mainly from the problem that, as Baez notes, „[w]hile Penrose originally intended for spin networks to describe the geometry of space-time, they are really better for describing the geometry of space.“ ([2], p.2) The dynamics (entering via the Hamiltonian constraint) is practically absent. So that the necessity arises to introduce the „evolutionary“ aspect in terms of some path integral formalism or of something which „sums over states“ (the latter given by means of spin networks). This refers mainly to the work of Rovelli and Smolin [4] who for the first time replaced the original spin networks by „embedded spin networks“, i.e. usual spin networks plus an immersion of their graphs into a three-dimensional manifold, calling equivalence classes of spin networks under diffeomorphisms *s-knots*. These do not live on a manifold, i.e. they are not located somewhere, but they actually *are* this somewhere. Since spin network states of type $\langle S|$ span the loop state space, one can see that any ket state $|\psi\rangle$ is uniquely determined by the values of the $\langle S|$ functionals on it:

$$\psi(S) := \langle S|\psi\rangle.$$

In this sense, a quantum state of the geometry of space is being described by a linear combination of spin networks. Areas and volumes take on a discrete spectrum of quantized values. And the transition amplitudes between states are computed as sums over spin foams. Hence, spin network states show up as elementary quantum excitations of the gravitational field carrying quanta of volume at their vertices (nodes) and quanta of area on their edges (links).

There is however (beside technical aspects), a conceptual problem to this: The point is that as loop quantum gravity is primarily based on canonical quantization, a concept of „fixed background time“ is necessary in order to describe a state (of the geometry). This means that it is not so clear what a „generic slice“ (of a spin foam) should actually be. In other words: What would be a „natural time distance“ or „lapse“ between spin network states? [5] On the other hand, this touches a more critical question, namely: „How fundamental is ‚fundamental‘?“ As it turns out, the expression „fundamental“ is utilized differently, in physics, and in philosophy. This can be seen as follows:

As quoted already, Baez refers to Penrose stating that he „intended for spin networks *to describe* the geometry of space-time.“ And he explicitly formulates:

„Spin networks were first introduced by Penrose as a radical, *purely combinatorial description of the geometry of spacetime*." ([2], pp.1-2, emphasis mine) As it appears, this is not quite true: Because Penrose derives his original idea from the fact that physics plays in two different „worlds", the continuous space of general relativity (a Lorentz manifold of dim = 4 and sign = – 2) on the one hand, and the space of quantum theory (a Hilbert manifold of high (or infinite) dimension and positive-definite sign) on the other. Assuming that a unification of both *within* the (physical) world would not be possible, his idea is to „build up both space-time and quantum mechanics simultaneously – from combinatorial principles – ..." ([1], p.151) Hence, what he does is to actually *leave* the physical world in a conceptual sense in order to start from an abstract structure from which spacetime could be eventually *derived*. In other words: This structure takes the (philosophical) connotation of a *substance* in the traditional metaphysical sense, referring to something which is beyond the world (beyond space and time that is) while spacetime as it is topic of both the theory of general relativity and of quantum theory is visualized under two different aspects of the same underlying entity, hence as *attributes* of that substance. The difficulty of actually performing this task of describing such an abstract structure (be it in mathematical, physical, or philosophical terms) is in the fact that space and time cannot be excluded from this conceptual discussion from the outset, because of epistemological reasons: we use them as the very *categories of thinking*. There is no way to express what we think otherwise than by means of (propositional) language. And languages (at least those of European descent) utilize explicitly space and time categories from which they derive their lexicology, syntax, and semantics. Hence, the conceptual emergence of 3-space as described in this early paper by Penrose refers to the *process of thinking*, i.e. to emergence in epistemological terms, rather than to a physical process, i.e. to emergence in ontological terms. The point is that we have to model the emergence of spacetime according to our available language (i.e. modaliter), while „in reality" (i.e. realiter), the combinatorial structure as introduced by Penrose is permanently *underlying* all what there is. In this sense, it is referred to as being *fundamental*. Therefore, general relativity actually describes spacetime as we can visualize it in classical terms when modelling it according to Einstein's theory based on empirical (i.e. observable and perceivable) aspects of the physical world. But spin networks as a combinatorial, abstract structure, serve to visualize an underlying entity we cannot actually perceive at all. In a (meta-theoretical) way, general relativity is a *concrete* model of the Universe, spin network theory is an *abstract* model of the *foundation* of the Universe. The difference between the former and quantum theory is simply that in the case of quantum theory, it is not so easy to differ between its empirically accessible parts and its purely interpretational parts. It is comparatively easy to recognize this conceptual problem as the ancient problem of the relationship between substance and attribute. [6]

This means that the original question („How fundamental is ‚fundamental'?") points to a conceptual (and thus philosophical) gap between the physical world and its foundation: Indeed, within the recent approach, spin networks serve to quantize space and determine the „boundary“ of the Planck scale. When discussing flux lines of a gravitational field piercing black hole horizons and exciting curvature degrees of freedom on the surfaces, Ashtekar and Krasnov can visualize each micro-state as an elementary bit of information. [7] But if there actually *are* quanta of space (and time), then there *are* space and time. (And not their foundations.) And if these quanta refer to bits, they refer to classical information, not to quantum information. In fact, Wheeler's conception of „It from Bit“ (mentioned by Ashtekar and Krasnov) has to be replaced now by a more subtle conception of „It from Qubit“. ([8], p.47)

But, even if so, the question is also whether it is possible to continue modelling the foundation (i.e. the truly *fundamental level* of physics) by utilizing classical concepts such as space, time, and even causality (indicating proper light cone structures). Because this would be then something *outside* the Universe. (As Schelling once formulated a long time ago: „ ... foundation is against that to which it is foundation, *non-being*.“ [9], p.440) This obviously speaks against a path integral approach, at least on the really fundamental level. A similar argument has been formulated earlier by Crane. [10] Therefore, the more recent approaches by Barrett and Crane put the emphasis on algebraic and categorial aspects of the theory, and mainly on topological quantum field theories. [11] On the other hand, very rarely, the substance-attribute relationship is even mentioned when talking about quantum gravity, but then only in passing: „We harbour no illusion that such traditional philosophical ideas are likely to be heuristically helpful, let alone a prerequisite, for theory-construction.“ ([12], p.40) Strong words indeed! Indicating that everything would have been tried before, and one would really know about substance. But this is not quite the case, because the knowledge among physicists – and analytical philosophers – of modern interpretations of Spinoza's substance metaphysics is not only limited, but practically absent, as can be empirically demonstrated even among those (few) who deal seriously with metaphysics at all. Even the relationship between Leibnizian and Spinozist theory is not well understood. Hence, what we can state is the following:

All of this uncovers the fact that the term „fundamental“ is usually utilized for the „boundary“ region of spacetime rather than for the truly fundamental level of consideration. Within the picture of loop quantum gravity, this boundary region is constituted by means of spin networks which define the „lowest level“ of (quanta of) space. Re-reading a famous passage of Rovelli's can clarify what is being meant with „fundamental“: There is an obvious contradiction between the proposition that „[w]e know that there is a regime in which all present *fundamental* theories break down“, and the proposition that „[i]n a sense, ‚there is no time‘ at the *fundamental* level.“ ([13], pp.207, 213) Because in the former case, there are no fundamental theories in this sense: It may be (or rather: it is likely) that Einstein has visualized his theory as fundamental when thinking of geometry as being substance, but the advent of quantum theory has destroyed this ho-pe. ([6], reference 6) On the other hand, quantum theory for itself is not fundamental enough, as the role of gravitation within attempted TOEs seems to indicate. Note that most approaches in quantum theory deal with an interpretation of the theory in its traditional form as it has been handed down to us from the times of Schroedinger and Heisenberg. There is hardly any reference in this as to the role of gravity. On the really fundamental level however, there is no space and no time at all, be it quantized or not. So the real problem is *to think* the foundation of the world in a way which marks clearly its categorial difference from the world itself, and then to re-construct the transition from this fundamental level to the boundary of (quantized) world.

Such an approach is the only possibility to actually avoid the „ontological vagueness" Goldstein and Teufel noted recently. ([14], p.276) They actually discuss the Bohmian approach. But their statement, that „[t]he unobserved physical reality becomes drastically different from the observed, even on the macroscopic level of daily life" ([14], p.281), originally thought of as criticizing „classical" quantum theory, takes on a completely new connotation now, if visualized within the framework of what we have discussed so far. Because indeed, there would be two practically disjoint regions now: the (physical) domain of what can be observed in principle, and the (fundamental) domain of what *cannot* be observed in principle. This is not quite in the pure tradition of physics, but we would not like to join in the stating of a theory being „scientific" dependent on the question whether there are empirical data available or not, contrary to current opinion. (Cf. [12], p.36) The point is simply that any TOE has to cover parts of reality (i.e. has to talk about them in propositional terms) which are not accessible to measurement in principle. Nevertheless, one would call it a scientific theory. This is not really a new insight. (Cf. [13], p.191) But this is also the place where modern philosophy actually enters the discussion. Because it is its genuine task to actually deal with this transition between foundation and world (possibility and actuality). [15] So instead of simply appealing all the time to philosophy ([13], p.182), or of listing problems that might be topics of philosophy ([14], p. 279sqq.), physicists should by now take some advice from a field which has cultivated consistent, systematic, and logical, if not mathematical thinking for a long time. In fact, it can be shown that quite a long time ago, within the period of Italian Renaissance, the introduction of mathematical argumentation into science, explicitly cultivated by physicists nowadays, corresponds to a somewhat clandestine re-introduction of previously criticized substance into the constitution of world views. ([16], p.181sq.)

So there are in fact precise though heuristic results as offered by philosophical insight: One point is the absence of causality on the truly fundamental level of physics. Another point is the possible relevance as to the explicit difference of the Leibnizian model (favoured by present physics as far as it is interested in philosophical conceptualization) as compared to the Spinozist approach.

As to the first we have to ask whether the re-conceptualization of spin networks in the sense of Penrose is not unwillingly falling back behind the original com-binatorial insight gained: namely to derive spacetime from a purely combinatorial structure. Smolin admits that spin network states as they are introduced now „are not purely combinatorial objects“ ([17], p.7), they actually show up as „a mixture of combinatorial spin networks, *causal sets*, and critical behaviour models.“ ([18], p.3, emphasis mine) Hence, the second is being „put in by hand“, the third is referring to self-organized criticality in the sense of the Santa Fe school (which might turn out to be independent of a substratum). This strategy of building in causality from the outset is put forward in a large number of papers. [19] However, the problem is in the transition from combinatorial to spatial and temporal structures: The causal sets showing up in the functor Past refer to events which are ill-defined on the fundamental level. (Without space and time one would not know what an event actually is.) And the causal history is defined as a succession (or even recursion) of Pachner moves which are essentially local unitary operators. Hence, the formulation of an amplitude for going from network 1 to network 2 in the well-known form of

$$A(1\rightarrow 2) = \sum (\text{histories } 1\rightarrow 2)\ \Pi\ (\text{moves per history})\ A_{\text{move}}$$

as a sum-over-histories. But this approach appears to be somewhat unsatisfactory, because we essentially talk about *combinatorial* moves. And the question „How much time needs a Pachner move?“ is probably as difficult to answer as the question „How much time needs a Reidemeister move?“. [20] The problem is similar to that of spin foams when talking about a „generic slice“ and the appropriate „lapse“. If the „time step“ is in the tetrahedra of the triangulating construction [21], can this then be rephrased in terms of some $t_p$ which generically accompanies $l_p$? This is not the idea of Penrose though: „The final theory that emerges must have a fundamentally non-local character.“ ([22], p.424) And we can add: „ ... must be *a-temporal* therefore.“

The other point is a little more involved, but basically on the same line of argument: The main difference between the Leibnizian and Spinozist concepts of space is that Leibniz explains away space and ends up with objects only such that space shows up as a cognitive approximation of the set of relationships among them. Spinoza on the other hand, explains away objects and stays with the spatial region alone. For him, objects show up as a cognitive approximation of the fact that some spatial region is *object-like* (we observe a pebble, because space is *pebbly* in some region). [6] In terms of Einstein’s diffeomorphism invariance, these two views seem to be equivalent (if actually visualizing space as a set of relationships anyway). However, for Spinoza, space as we know it was essentially *the only attribute* of substance. Or more precisely: What humans perceive of space (and call physical space) is what they are capable of perceiving *under the only attribute* (res extensa) which falls into their mode of being. The other attribute (res cogitans) gains by this the connotation of a materialistic derivation, and time shows up as its property. Obviously, what humans perceive of the attributes cannot be fundamental.

Both these points, causality being visualized as an emergent rather than a fundamental concept, and the Spinozist view, spacetime being *self-relational* rather than relational, challenge Smolin's criteria for a background independent approach (namely numbers 3, 5 and 6, respectively). [23]
If we are visualizing therefore space quanta in terms of spin network states and include an intrinsic ordering principle called *time* (evolution), the largest common denominator of this consideration might be to view the web of spin networks as the „boundary layer" of worldliness, as the borderline between the fundamental and the empirical. Hence, it is natural to ask two questions first: What would come „below" this boundary level of the world (in terms of spin networks), and what would come „above" this level? Some aspects of these questions are being discussed in the next two sections.

## 2 Downward Causation: Quantum Computation

Note first that when we speak of „causation" we refer to the consistent unfolding of our ideas dealing with the re-constructing of hierarchical levels of reasoning rather than with concrete physical processes. The point is that the quantum legacy stands somewhat against this acquired consistency, because its characteristic aspects tend to counteract rational assumptions. This can be clearly seen with respect to the concept of time: Recall Feynman's famous „integral argument" in favour of visualizing the Universe as a quantum computer. Then, the integral of the celebrated form,

$$P(A \to B) = \left| \int_\gamma \exp\,(iS/\hbar)\, d\Gamma \right|^2$$

can be essentially interpreted such that it defines the probability for a physical system going from some state A into another state B. The idea is that this transition is actually being performed over all possible paths $\gamma$ *at the same time* – but that we can observe only that path for which $\delta S = 0$, and we call this the „classical path". However, the action here is simply defined in terms of the usual time integral of the appropriate Lagrangian,

$$S = \int L \; dt,$$

and we wonder what the usual time is actually meaning on this fundamental level, where transitions take many ways at the same time. (This problem is of course the same for utilizing the time-dependent Schroedinger equation even if talking of unitary transformations. Indirectly, this also causes the interpretational difficulties of the Hamiltonian constraint.) ([14], p.279)

On the other hand, leaving the classical time parameter aside for a moment, the „computer paradigm“ is plausible in the first place, provided some discrete computing algorithm can be found of which space and time turn out to be the classical approximation. Intuitively, it is the spin networks themselves that can be offered as a fundamental sort of „information channel“, their web structure constituting a „cosmic internet“ of very basic kind, possibly in the recently proposed sense of Zizzi. [24] There is an alternative approach to computing networks of this kind by Hitchcock. [25] In principle, this universal computer can be visualized then as analogy of a system which percolates information through available sites of a cellular (automaton) structure. This sort of model has been mentioned several times before. [26]
But there is another point to that: In the approach of Barrett and Crane, the idea is to generalize topological state sum models in passing from three to four dimensions by replacing the characteristic SO(3) group with SO(4), or its appropriate spin covering, SU(2) x SU(2), respectively. [11] The concept of spin networks is also generalized then, by introducing graphs with edges labelled by non-negative real numbers (called „relativistic spin networks“). Applying this kind of „spin foam“ model to Lorentzian state sums demonstrates their finiteness in turn implying a number of choices made from physical and/or geometrical arguments. [27] The really interesting aspect of this is its relation to the group SL(2,C): because this is the double cover of SO(3,1) and the complexification of SU(2) which in turn is the double cover of SO(3). On the other hand, SL(2,C) is the group of linear transformations of $C^2$ that preserve the volume form. Thanks to an e-mail crash course on these matters referring to the Barrett-Crane model and made available online by John Baez and Dan Christensen [28], where they use the terminology of the former’s quantum gravity seminar [29], it is easy to understand that constructions in the sense of Barrett-Crane turn out to be invariant under SL(2,C). In other words, we essentially deal with states in $C^2$ which are spinors. And it is from quantum theory and special relativity that we know about their relevance. [30] On the other hand, as Baez notes, and as we will see shortly, a state in $C^2$ can also be called a *qubit*. So „[w]hat we [a]re really doing, from the latter viewpoint, is writing down ‚quantum logic gates‘ which manipulate *qubits* in an SU(2)-invariant [in fact, SL(2,C)-invariant] way. We [a]re seeing how to build little Lorentz-invariant quantum computers. From this viewpoint, what the Barrett-Crane model does is to build a theory of quantum gravity out of these little Planck-scale quantum computers.“ ([28], p.42) This is obviously very much on line with the arguments of Zizzi, Hitchcock, Lloyd and others. ([8], [24] – [26]) Moreover, it is referring to the explicit level of spin networks: That is, the aforementioned „boundary layer“ between the physical world and its foundation shows up as a „shift of quantum computing“ processsing the fundamental information necessary for performing the transition from foundation to world (or in other words: for actually *producing a world* out of its foundation).

Note however that consequently, when we are dealing with quantum computation, we are still talking about this boundary layer, not yet about the foundation itself. Because what we encounter when talking about time, we also encounter when thinking of linearity: the quantum legacy again. This means that the computational base pair leads naturally to a definition of a qubit state as a vector in a two-dimensional complex vector space with inner product such that a normalized vector can be represented by an expression of the form

$$\alpha|0> + \beta|1>; \ \alpha, \beta \in C,$$

such that

$$|\alpha|^2 + |\beta|^2 = 1.$$

The collection of n such qubits is a quantum register (of size n); and a quantum logic gate is simply a device which performs a fixed unitary operation on selected qubits in a fixed period of time. Hence, a quantum network consists of quantum logic gates whose computational steps are being synchronized in time (parallel). The size of the network is equal to the number of gates. Again, what we see is that the problem is „in the time" showing up here within the usual framework of qantum theory. So when looking for an „underlying fundamental level" beyond space and time, what we have to do is to step beyond this framework. It will not really help to pass to „entangled qubit states" (i.e. those which are non-separable and cannot be written as a tensor product), so long as we define quantum computers in terms of a family of quantum networks performing the activity of quantum computation, and visualizing the latter as a *unitary evolution* of these networks. [31] In principle, this context is not being left in all of the possible applications of entanglement, be it of maximal or mixed-state type. ([32], pp.142, 222) In the end, disentanglement shows up then as an intermediate state on the way from entanglement towards classicity, but it is not anything beyond the physicality of the aforementioned „boundary layer": The use of a background time analogue (by means of unitary transformations) precludes a real extension on to a truly fundamental level of modelling.

Another question would be whether the „direction of development", namely from entanglement towards disentanglement is actually the correct choice of describing the onset of classicity. Because, if we remember the phenomenon of decoherence as a means to produce classicity, then this is basically the result of entanglement (of qubits with some environment) rather than disentanglement:

$$(\alpha|0> + \beta|1>)|m> \rightarrow \alpha|0>|m_0> + \beta|1>|m_1>$$

such that the density operators are of the form

$$\begin{pmatrix} |\alpha|^2 & \alpha\beta^* \\ \alpha^*\beta & |\beta|^2 \end{pmatrix} \rightarrow \begin{pmatrix} |\alpha|^2 & \alpha\beta^* <m_0|m_1> \\ \alpha^*\beta <m_1|m_0> & |\beta|^2 \end{pmatrix},$$

and as $<m_1|m_0> \rightarrow 0$, „coherences" vanish. A similar transition we would need from the foundation onto the „boundary layer" of (quantum) spin networks. If we visualize the foundation as a structure without any spatial and temporal connotation, but would like to stay with a computational analogue, then we would think of this sort of „fundamental computational process" as some kind of fluctuation permanently exchanging spin network states (which themselves serve as quantum logic gates for quantum computation „on the other side" of the boundary layer configuration). But unlike the idea of evolving spin networks in a spin foam fashion according to some intrinsic time measure defining the generic slices between spin networks, this fluctuation would be essentially *non-local*, in the sense that as seen under the perspective of the foundation, it would be always and everywhere. (This has been earlier referred to as „intrinsic vibrational states" [26], third reference.) If visualized under the classical perspective of a physical observer within the world however, this fact could only be consistently conceptualized, if thought of in terms of differentiating spatial and temporal distances, and bringing representations of this kind into some sort of digital succession. The difficulty of rephrasing such a model utilizing a language which is basically structured in a spatial and temporal manner, is topic of presently ongoing work. (Cf. [5], first reference) Originally ([26], third reference), it was thought of utilizing recent insight of Bieberich for this task. [33] This looks still very promising and is part of the aforementioned project under way.

## 3 Upward Causation: Wilson Loops & Knots

The other „direction" of the argument is easier to tackle. This is mainly so because loops and knots have been discussed in some detail by now. Note that according to the standard terminology (we follow here essentially [34]), a loop in some space $\Sigma$, say, is a continuous map $\gamma$ from the unit interval into $\Sigma$ such that $\gamma(0) = \gamma(1)$. The set of all such maps will be denoted by $\Omega\Sigma$, the loop space of $\Sigma$. Given a loop element $\gamma$, and a space $\mathcal{A}$ of connections, we can define a complex function on $\mathcal{A} \times \Omega\Sigma$, the socalled *Wilson loop*, such that

$$T_A(\gamma) := (1/N)\, \mathrm{Tr}_R\, P \exp \int_\gamma A.$$

Here, the path-ordered exponential of the connection $A \in \mathcal{A}$, along the loop $\gamma$, is also known as the holonomy of A along $\gamma$. The holonomy measures the change undergone by an internal vector when parallel transported along $\gamma$. The trace is taken in the representation R of G (which is the Lie group of Yang-Mills theory), N being the dimensionality of this representation. The quantity measures therefore the curvature (or field strength) in a gauge-invariant way. For gravity, it is more appropriate to choose Wilson loops that are constant along the orbits of the diffeomorphism action so that

$$T(\varphi \circ \gamma) = T(\gamma) \text{ for all } \varphi \in \mathrm{Diff}(\Sigma).$$

Note that because the original idea of this came from Witten who introduced a new class of manifold invariants in terms of generalized Feynman integrals, namely of the form,

$$Z(M) = \int dA \exp[(ik/4\pi) S(M, A)],$$

with M being a 3-manifold without boundary, A the gauge field (connection) on M, and the action S the integral over M of the trace of the Chern-Simons 3-form, we are on the „safe side" here with the path integral formalism, because we are dealing with the level „above" the aforementioned „boundary layer" as represented by the spin networks themselves. This can be seen as follows: Over a given loop $\gamma$, the expectation value $< T(\gamma)>$ turns out to be equal to a knot invariant (the „Kauffman bracket") such that when applied to spin networks, the latter shows up as a deformation of Penrose's value $V(\Gamma)$. This is mainly due to the fact that

$$< T(\gamma)> = K^k(\gamma) = (1/Z) \int d\mu(A) \exp [(ik/4\pi) S] T(\gamma, A).$$

So, for any spin network $\Gamma$ (replace $\gamma$ by $\Gamma$), the old relation holds up to regularization. Hence, spin networks are deformed into quantum spin networks (which are essentially given by a family of deformations of the original networks of Penrose labelled by a deformation parameter $q = \exp (4\pi/(k+2))$ for the Chern-Simons case). Note also that the Chern-Simons invariant is important when having a non-zero cosmological constant $\Lambda$, because there is an exact physical state of quantum gravity given by $\Psi_{CS} (A) = \exp (k/4\pi\ S^{CS} (A))$, where k is actually related to Newton's constant by $G^2\Lambda = 6\pi/k$. This state can be shown to reproduce $K^k(\Gamma)$ above.

As Kauffman has shown [35], the condition of being a knot invariant can be thought of as being equivalent to the diffeomorphism constraint. In particular, knot states, i.e. loop functionals that depend only on the knotting of the loops, can completely solve the diffeomorphism constraint. [36] But then, they are knotted states of spin networks rather than spin networks themselves, and they *cannot* be the true elementary quanta of space. ([36], p.16) They rather depend on loops as their independent variables. (And in fact, they have non-local dependence on the connection.) ([37], p.63) But as far as all information present in the holonomy can actually be reconstructed from the Wilson loops, the latter become fundamental variables in themselves. The point is however, that they constitute „the worldly side“ of increasing structural complexity. In other words: They are structurally more complex than (unknotted) spin networks. In this sense, loops can carry diffeomeorphism-invariant information which is not necessarily in their knot-theoretical representation. ([34], p.7) And they can be shown to solve the Wheeler-deWitt equation, in the first place. ([36], p.16; [38], p.435) But as to the path integral formalism, we are on the „safe side“, because above spin network level, we do not have any problems with the reasonable interpretation of time parameters or similar things.

## 4 TQFT: Cobordisms & Categories

From here, we can discuss the further unfolding of structure, parallel to results in topological quantum field theory (TQFT), as discussed for the simplified case of BF theory. [2] Be $\mathcal{A}_0$ the moduli space of flat connections on P, the physical phase space, and be $\mathcal{G}$ the group of gauge transformations of the bundle P. Recall then the definition of a *spin foam* which is very much alike the one for a spin network, only one dimension higher. Hence, a spin foam is essentially taking one spin network into another, of the form F: $\Psi \rightarrow \Psi'$. Just as spin networks are designed to merge the concepts of quantum state and geometry of space, spin foams shall serve the merging of concepts of quantum history and geometry of space-time. As we have already mentioned, very much like Feynman diagrams do, also spin foams can be used to evaluate information about the history of a transition of which the amplitude is being determined. Hence, if $\Psi$ and $\Psi'$ are spin networks with underlying graphs $\gamma$ and $\gamma'$, then any spin foam F: $\Psi \rightarrow \Psi'$ determines an operator from $L^2(\mathcal{A}_\gamma / \mathcal{G}_\gamma)$ to $L^2(\mathcal{A}_{\gamma'} / \mathcal{G}_{\gamma'})$ denoted by $\mathcal{O}$ such that

$$<\Phi', \mathcal{O}\, \Phi> = <\Phi', \Psi'><\Psi, \Phi>$$

for any states $\Phi$, $\Phi'$. The evolution operator Z(M) is a linear combination of these operators weighted by the amplitudes $Z(\mathcal{O})$. Obviously, we can define a category with spin networks as objects and spin foams as morphisms.

Hence, it turns out, in fact, that $L^2(\mathcal{A} / \mathcal{G})$ is actually being spanned by spin network states. Call such a state $\Psi \in$ Fun$(\mathcal{A} / \mathcal{G})$ so that any spin network in S ( =

space) defines such a function. Because Fun is an algebra (namely consisting of all functions on A of the form $\Psi(A) = f(T \exp \int_{\gamma 1} A \ldots T \exp \int_{\gamma n} A)$, where f is a continuous complex-valued function of finitely many holonomies which are represented here by the integral expressions), multiplication by $\Psi$ defines an operator on Fun. We call this operator *spin network observable*.

So what we essentially do is the following: Given the (n-1)-dimensional space S and a triangulation of S, choose a graph called the dual 1-skeleton. Express any state in Fun(A /G) as a linear combination of states coming from spin networks whose underlying graph is this dual 1-skeleton. Define now space-time as a compact oriented cobordism M: S $\rightarrow$ S', where S, S' are compact oriented manifolds of dimension n-1. Choose a triangulation of M such that the triangulations of S, S' with dual 1-skeletons $\gamma$, $\gamma$' can be determined. The basis for gauge-invariant Hilbert spaces is given by the respective spin networks. Then the evolution operator Z(M): $L^2(A_\gamma/G_\gamma) \rightarrow L^2(A_{\gamma'}/G_{\gamma'})$ determines transition amplitudes <$\Psi$', Z(M) $\Psi$> with $\Psi$, $\Psi$' being spin network states. Write the amplitude as a sum over spin foams from $\Psi$ to $\Psi$': $< , > = \sum_{F:\Psi\rightarrow\Psi'} Z(F)$ plus composition rules such that Z(F') o Z(F) = Z(F' o F). This is a discrete version of a path integral. Hence, re-arrangement of spin numbers on the „combinatorial level" is equivalent to an evolution of states in terms of Hilbert spaces in the „quantum picture" and effectively changes the topology of space on the „cobordism level". This can be understood as a kind of *manifold morphogenesis* in time: Visualize the n-dimensional manifold M (with $\partial M = S \cup S'$ - disjointly) as M: S $\rightarrow$ S', that is as a process (or as time) passing from S to S'. This is the mentioned *cobordism*. Note that composition of cobordisms holds and is associative, but *not commutative*. Consequently, these results can also be formulated in the language of category theory: As TQFT maps each manifold S representing space to a Hilbert space Z(S) and each cobordism M: S $\rightarrow$ S' representing space-time to an operator Z(M): Z(S) $\rightarrow$ Z(S') such that composition and identities are preserved, this means that TQFT is a functor Z: nCob $\rightarrow$ Hilb. Note that the non-commutativity of operators in quantum theory corresponds to non-commutativity of composing cobordisms, and adjoint operation in quantum theory turning an operator A: H $\rightarrow$ H' into A*: H' $\rightarrow$ H corresponds to the operation of reversing the roles of past and future in a cobordism M: S $\rightarrow$ S' obtaining M*: S' $\rightarrow$ S.

So what we have here in the end, is the structural mediation of the micro-level of physics with the macro-level of (physical) forms. As re-arrangements of spin numbers in the combinatorial network correspond to re-arrangements of spatial quantizations within the triangulation picture, and as both correspond to a macroscopic evolution of form in terms of spacetime as it can be observed by some classical observer, this can be visualized as the true onset of a theory of emergence (once, a generalization beyond BF theory is being achieved).

## 5 SOC: The Mediation of Micro & Macro

What TQFT is achieving for the structural mediation, self-organized criticality (SOC) in the sense of the Santa Fe school is doing for the dynamical mediation in terms of an explicit transition rule („fourth law stuff“). [39] Stuart Kauffman starts from the basic idea that the Universe is non-ergodic with respect to its hierarchical complexity. This is the underlying motivation for assuming the existence of a „fourth law“ of thermodynamics which actually steers the dynamics of evolutionary processes. The definition is essentially that a transition (of a given physical system) from one state to another is such that it leads into the „adjacent possible“ meaning the set of all possible states for that system which are one reaction step away from the already actualized states. Hence, compatible with the principles of self-organized criticality [40], the idea is to base global outcomes of processes on local interactions. Because this assumption implies also that the actual flow from the possible to the actual would be such that the dimensionality of the „adjacent possible“, on average, expands as rapidly as it can (probably constituting by this an „arrow of time“), the Universe altogether might tend to flow towards maximal complexity. With a view to percolation models, Kauffman has, in close cooperation with Smolin, applied these basic aspects to spin networks themselves. He visualizes them as knotted structures, and as such as combinatorial objects similar to symbol strings of a grammar, becoming „collectively autocatalytic“, namely as „knots acting on knots to create knots in rich coupled cycles not unlike a metabolism.“ ([39], ch.7, p.2) We will discuss these aspects again in the second part of this present paper. More important here for the time being is to notice the basic consequence of such an approach:

When we try to re-construct the mediation between the micro-level and the macro-level of the physical world as we can observe it, what we actually do is to re-construct our own modelling of the world. All what we can do is to model the world according to what we perceive of it. The set of such models we call knowledge. [6] At most, we can model the foundation of the world according to what we know about the world in this sense. Hence, we proceed in a somewhat *transcendental* fashion: We permanently ask for the necessary conditions under which the world exists, given the facts we know about it. But we know beforehand that the outcome of this must be necessarily incomplete, simply because our means of perception are incomplete. (They are even incomplete with respect to what we actually know.) [42]

This basically implies that in principle, there is no chance to eventually complete the information we can extract about the world altogether. Even worse: According to our own models we can well assume that the lifetime of humans is finite as compared to the lifetime of the Universe. Hence, there is not even the chance to eventually develop an all-encompassing „final“ model of our own component of the world. As I have shown elsewhere in some detail [43], this has significant ethical implications. (We come to that in the second part of this pre-sent paper.) Nevertheless, this incomplete situation for human life is an intrinsic part of the whole process. It could not be otherwise. Therefore, what humans do all the time (namely to model their world) has to be visualized as a generic phenomenon of the very process they actually try to model. This is the reason for a permanent self-reference which is being built in from the beginning into all the modelling as it is achieved in terms of the various forms of research. In other words: Developing a structural mediation between micro- and macro-level, and finding a dynamical rule which generates it, is not only performing the required re-construction of a permanently becoming knowledge as part of the process of which it actually *is* knowledge, in the first place, – and communicating it into the social collective by means of appropriate semiological techniques –, it also is the very process *itself*. But what we call „process“ in worldly terms, is not a process at all, if visualized under the perspective of the foundation of the world. Because in order to define a process we need the adequate categories within whose framework to phrase the definition. And for us, these are space and time. So, in the end, we are back to the beginning, and we note the following:

The most we can do is to assume that *there actually is something* at all, independent of human existence. This is a basic ontological statement, and we call it a statement of „realism.“ But *what* that something actually is, we cannot know. We can only model aspects of it. This is the related epistemological statement. What we have to understand is that these statements do not primarily refer to our incomplete knowledge, but that instead it is the very tension between the „That“ and the „What“ of existence, which produces our framework of categories, in the first place. What does that mean in practical terms? It means that what we can do is to model the world according to the insight gained by developing our ruling paradigms: A number of centuries ago, it was appropriate to model the Universe according to the clock paradigm, because the mechanism of the clock (a first digitalization of some process, by the way) served as an innovative model for describing organizing processes. Obviously, they were not self-organized at the time, the existence of some maker of the universal clock being assumed in the first place. Nowadays, it is more appropriate to model the Universe according to the computer paradigm, because it is this technology which has replaced the classical (and mechanical) technology of a clockwork. And the universal computer is thought of as being initiated by self-assembly. Hence, the Universe is visualized as being self-organizing.

Basically therefore, modern substance, as we have seen earlier, can be modelled according to a universal (quantum) computer. (Provided we do not forget that this sort of metaphore is referring to the boundary of substance rather.) The mediation from spin networks via knotted structures up to macroscopic structures as performed for a simplified case in terms of topological quantum field theory, turns out as a specific re-construction fitted to the original paradigm. The same is true when looking for a dynamical rule which is steering the explicit unfolding of structure. As we will see shortly, we can recover the same aspect again, when talking about life in the Universe. In fact, the combination of the computer paradigm with biological selection is the best we can achieve for the present modelling of the world.

## 6 The Cosmic Selection of Life

In this sense, Smolin has introduced a generalized concept of selection by talking about selection of Universes rather than of species. [26] This cosmic selection rule can be interpreted as the worldly version of the aforementioned „knot metabolism" as discussed by Kauffman. The biological species definition is being replaced by a type definition of Universes.
Hence, Smolin postulates that Black Holes, instead of representing singular points of „physical breakdown" in the Universe, serve as a sort of tunnel into the early states of a novel Universe being actually produced by a „bouncing" process under extremal conditions, mapping the micro-version of a „final singularity" to the „initial singularity" of a Universe coming to life. This is a conception similar to Wormholes and/or Baby-Universes. The idea is then that the production of Universes is such that those with a high production rate of Black Holes (and thus of novel Universes) are selected. Consequently, Smolin speaks of a theory of „cosmological natural selection". ([26], pp.88, 96, 108; [41]) Although this approach is not free of misunderstanding and drawback (I have discussed this in some detail in [26], second reference), its advantage is first, the utilization of aspects of Waddington's „fitness landscape" deriving primarily from the „galactic ecology" of interstellar matter, and second, its compatibility with the principles of self-organized criticality. ([26], pp.159, 168-171) This also sheds some more light onto the position which life itself is taking within this picture: „Life perhaps might be seen to have evolved a way to ride these flows and cycles [of self-organized criticality] the way a surfer rides the flow of energy in water waves." ([26], p.154sq.)
In the meantime, it has been shown that there are new arguments in favour of a biological analogue to selection: Patel has argued that the relevance of Grover's search algorithm for the information processing in genetic DNA reveals that life has taken the route of digitalizing its information. [44] Based on this, Zizzi assumes that the observation probability of a state, when applying amplitudes in the quantum algorithm approach, can be related to a similar value which is valid for the early Universe, provided a suitable de Sitter model is being chosen. [45] This also sheds some more light on the question whether it would be useful to generalize the concepts of life and/or consciousness with a view to their universal role within the world:
The point is here that these concepts fall under the „rule of generality": That is, if we assume that the Universe altogether is a living entity, then finally, we end up with having no reasonable meaning left for the concept of life. (If everything is life, nothing is.) If, on the contrary, we would like to save the differentiating function of the concept, we cannot possibly implant life and/or consciousness (as we know it) on the quantum level of the consideration. This is mainly so, because the concepts we utilize (by means of our language) can be shown to be emergent concepts. If so, thinking itself is emergent, and so is consciousness, and finally, life altogether. On the other hand, the necessary conditions for eventually having life at all, have to be intrinsic conditions of the process itself. From what we have discussed before, we know already that „process“ carries a model connotation only rather than being of ontological significance. Hence, what we do is to utilize concepts of life and consciousness in order to symbolize the self-reference we experience when producing models of the world such that each model contains at least one self-model. But because there is not really a process, the underlying correlate of what we call life (or consciousness) must be actually present on the truly fundamental level all the time. We can call this correlate „proto-consciousness“ (or „proto-life“ as for that), but we cannot overcome its merely epistemological connotation in really ontological terms. And this will be the starting point for the second part of this present paper.

## Preliminary Conclusion

What we have found out so far is that, so long as we stay within the field of physics proper, presently available models of loop quantum gravity are able to provide for a theoretical basis to eventually re-construct the evolution of the world in terms of some theory of emergence, mediating the world as it is being observed with its foundation. Given certain ontological qualifications, this approach turns out as a suitable case study for actually differentiating between world and foundation in scientific terms. At the same time however, it becomes quite clear that until now, this significant difference has not yet been topic of physical theory. Hence, the concept of „fundamental“ has to rephrased somewhat; and the heuristic implications of its new utilization have to be discussed in some detail. This is where modern philosophy comes in. Contrary to the sciences which take a specialized perspective as to their respective field, philosophy tries to achieve a reasonable totalization of all the scientific fields. In this sense, it is generalistic rather than specialistic. Only the totalization of the world, not the mere summing of all its specialized perspectives, can relate it properly to its foundation. And it is only this relationship which can serve as a universal framework of orientation for our wordly existence.

## Acknowledgements

For clarifying and illuminating discussions in an early stage of this paper while meeting at the icmp2000 at Imperial College London I thank Paola Zizzi (Padova) and David Robson (London/Cambridge).